\documentclass[aps,prl,10pt,twocolumn,groupedaddress]{revtex4-1}

\usepackage{graphicx}
\usepackage{hyperref}
\usepackage{amsmath}
\usepackage{amsfonts}
\usepackage{amssymb}
\usepackage{subfigure}
\usepackage{times}
\begin{document}

%\preprint{}

\title{Origins of nonlocality near the neutrality point in graphene}


\author{Julien Renard}
%\email{renardj@phas.ubc.ca}
%\affiliation{Department of Physics and Astronomy, University of British Columbia, Vancouver, British Columbia, V6T1Z4, Canada}
\author{Matthias Studer}
%\affiliation{Department of Physics and Astronomy, University of British Columbia, Vancouver, British Columbia, V6T1Z4, Canada}
\author{Joshua A. Folk}
\email{jfolk@phas.ubc.ca}
\affiliation{Department of Physics and Astronomy, University of British Columbia, Vancouver, British Columbia, V6T1Z1, Canada}
%%%%{Your e-mail address}
%\thanks{}

\date{\today}

\begin{abstract}

We present an experimental study of nonlocal electrical signals near the Dirac point in graphene. The in-plane magnetic field dependence of the nonlocal signal confirms the role of spin in this effect, as expected from recent predictions of Zeeman spin Hall effect in graphene, but our experiments show that thermo-magneto-electric effects also contribute to nonlocality, and the effect is sometimes stronger than that due to spin. Thermal effects are seen to be very sensitive to sample details that do not influence other transport parameters. 

\end{abstract}

%\keywords{}

\maketitle

Nonlocality in an electronic device typically refers to the appearance of a voltage across contacts that are well outside the path one might expect an excitation (charge) current to follow.  One way for nonlocal voltages to arise is when the excitation current path is significantly modified from what would be expected by ohmic considerations. In the quantum Hall regime, for example,  current is carried around the edge of a sample while the bulk is insulating \cite{Mce90}.  Another common source of nonlocality is heat or spin currents that may be induced in a sample by charge excitation, but flow in directions not aligned with the exciting electric field \cite{joh85,bak10,zue09}.  Because nonlocality is in general associated with  nontrivial electronic interactions in the sample, nonlocal measurements are a powerful tool for investigating these interactions in novel materials.

It was recently pointed out that extreme levels of nonlocality are observed in charge-neutral graphene ($n$=0) subject to a large out-of-plane magnetic field \cite{aba11b}.  Although one might immediately attribute this phenomenon to edge state transport in the quantum Hall (QH) regime, the effect was observed at temperatures far above where QH effects disappear; furthermore, edge channels are not expected at the charge neutrality point, except in ultra-high mobility samples showing broken symmetry states.  Instead, it was argued that spin currents were responsible for the observed nonlocality \cite{aba11b,aba11}.

The prospect of generating large spin currents in graphene is exciting both for scientific and technological applications.    From a technological point of view, pure spin currents may form the basis for a new generation of  devices with much lower power consumption.  Graphene is especially promising for future spintronic technologies due to the material's weak intrinsic spin-orbit interaction.  From a scientific point of view, large discrepancies remain between theoretical predicted and experimentally measured spin relaxation times in graphene \cite{ert09,tom07,han10,han11}.  More powerful tools to generate and measure spin currents in graphene may help bring the two together.

In this paper, we investigate the origins of nonlocality in graphene, with an aim of quantifying the contribution of spin effects.  Nonlocal voltages are found to vary from sample to sample, even from region to region within the same sample.  More importantly, the fraction of the signal due to spin also varies from nearly 100\% down to negligible contributions.  We show that thermoelectric contributions to the signal are often as large or larger than the part due to spin, resulting from the Nernst effect and its inverse, the Nernst-Ettingshausen effect \cite{zue09,che09}.

\begin{figure}
\subfigure{\label{fig1a}}
\subfigure{\label{fig1b}}
\includegraphics{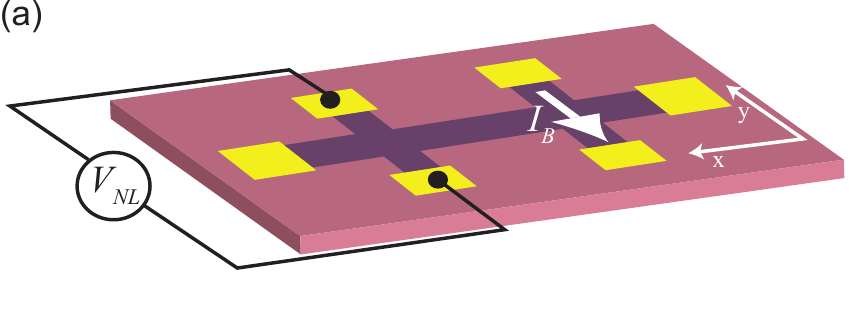}
\includegraphics{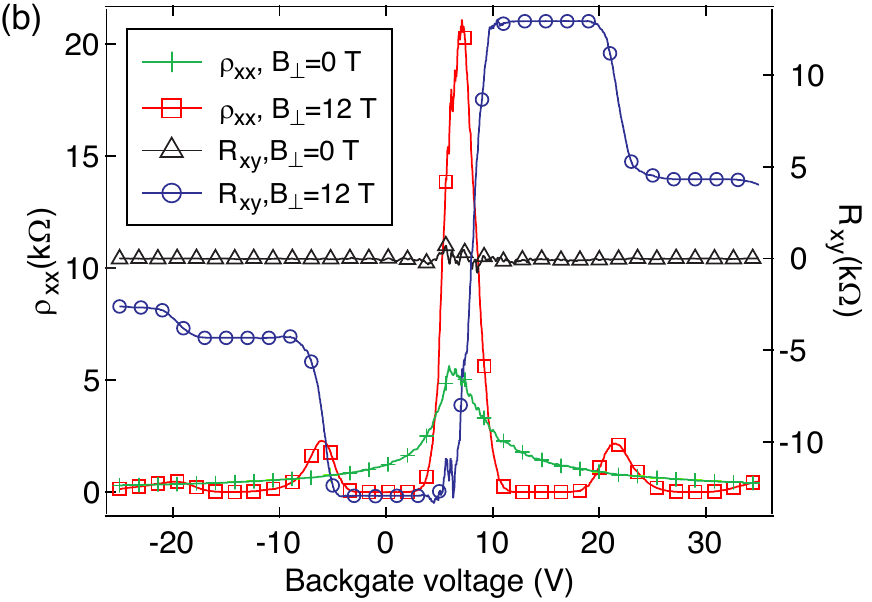}
\caption{\label{fig1} \subref{fig1a}: geometry of the experiment. A current $I_B$ is passed on one side of the graphene Hall bar and the voltage $V_{\rm NL}$ is measured in a nonlocal manner on the other side to give the nonlocal resistance: $R_{\rm NL}=\frac{V_{\rm NL}}{I_B}$. \subref{fig1b}: Local transport measurements of $\rho_{xx}$ and R$_{xy}$ at zero magnetic field and at 12T in the quantum Hall regime (Sample A, 4.2K).}
\end{figure}

The first nonlocal measurements in graphene were based on a conventional graphene Hall bar geometry similar to that shown in Fig.~\ref{fig1a} \cite{aba11b} .   A charge current, $I_{B}$, driven across the Hall bar between a pair of transverse contacts (the injector), generated a large nonlocal voltage, $V_{\rm NL}$, across a different pair of transverse contacts (the detector) at $n$=0 when the device was placed in a large out-of-plane magnetic field.  This observation was explained as arising from a ``Zeeman spin Hall effect" (ZSHE) and its inverse. This mechanism is different from the usual spin Hall effect (SHE) that appears in the presence of spin-orbit coupling, which in graphene can be induced by a weak hydrogenation \cite{Bal13}.

In the case of the ZSHE, Zeeman splitting due to magnetic field would induce spin-up electrons and spin-down holes at $n$=0 due to graphene's gapless band structure.  When a charge current is driven between the injector contacts (y-direction, Fig.~\ref{fig1a}), the magnetic field's orbital effect (the classical Hall effect) would split electron and hole states, driving a spin current perpendicular to the charge current and therefore along the length of the Hall bar (x-direction, Fig.~\ref{fig1a}) \cite{aba11b, aba11}.  At the detector, the spin current along the Hall bar would have the inverse effect, inducing a transverse voltage across the detector contacts (y-direction, Fig.~\ref{fig1a}).

The ZSHE depends on both orbital and Zeeman effects of a magnetic field, that is, the phenomenon depends on both the out-of-plane component of the magnetic field, $B_\perp$, and the total field $B_{\rm tot}=\sqrt{B_\perp^2+B_\parallel^2}$.  Taking both the ZSHE and its inverse into account, the nonlocal resistance predicted for the effect described above is\cite{aba09,aba11,aba11b}: 

\begin{equation}
R_{\rm NL}\equiv\frac{dV_{\rm NL}}{dI_{B}}\propto \frac{1}{\rho_{xx}}(\frac{\partial \rho_{xy}}{\partial \mu}E_Z)^2,
\label{rnlequ}
\end{equation}

\noindent where $\mu$ is the chemical potential.  The longitudinal  and transverse resistivities, $\rho_{xx}(B_{\perp})$  and $\rho_{xy}(B_{\perp})$ respectively, depend weakly if at all on $B_\parallel$, but strongly on $B_\perp$ in ways that are often difficult to predict.  The Zeeman energy $E_Z$, on the other hand, is simply proportional to $B_{\rm tot}$, so ZSHE predicts $R_{\rm NL}=\beta(B_\perp) B_{\rm tot}^2$ where $\beta(B_\perp)$ depends on sample details and on $B_\perp$ but not on $B_\parallel$.  Changing $B_\parallel$ while leaving $B_\perp$ fixed therefore provides a simple test for spin contributions to $R_{\rm NL}$.

Three samples (A,B and C) were prepared from graphene exfoliated on SiO$_2$. Sample C was obtained by thermal cycling sample A (including an annealing step at 200C in N$_2$/H$_2$ forming gas) resulting effectively in another sample with a different disorder. Following thermal evaporation of metallic contacts (Cr 0.5nm/Au 100 nm) the samples were etched into Hall bars using oxygen plasma.  The carrier density was controlled using backgate voltage $V_{\rm BG}$.  Channel widths were 800-900 nm; the distance between the classical current path and the detector ranged from 2.7$\mu$m to 3.5$\mu$m (see Fig.~\ref{fig1a}).   Quantum Hall measurements confirmed monolayer character (Fig.~\ref{fig1b}), with mobilities in the range 5000-10000 cm$^2$/Vs at 4.2K.   All measurements presented in this work were taken using lock-in techniques at $f\leq$12Hz; frequencies were confirmed to be in the DC limit.

An AC bias current $I_{B}$ of 10's of nA was applied across the injector contacts while the nonlocal voltage was monitored either at the first harmonic, $V_{\rm NL}^f$, or second harmonic, $V_{\rm NL}^{2f}$, of the excitation, giving nonlocal differential resistances $R_{\rm NL}^{f}\equiv dV_{\rm NL}^{f}/dI_{B}$ and $R_{\rm NL}^{2f}\equiv dV_{\rm NL}^{2f}/dI_{B}$. The nonlocal voltages were measured using a high input impedance voltage preamplifier (1T$\Omega$ in DC) to ensure the measurement had no effect on the current path. Measurements were carried out in a variable temperature probe (4K-80K) with an Attocube rotator, in magnetic fields up to 12T. 

\begin{figure}
\subfigure{\label{fig2a}}
\subfigure{\label{fig2b}}
%\subfigure{\label{fig2c}}
\includegraphics{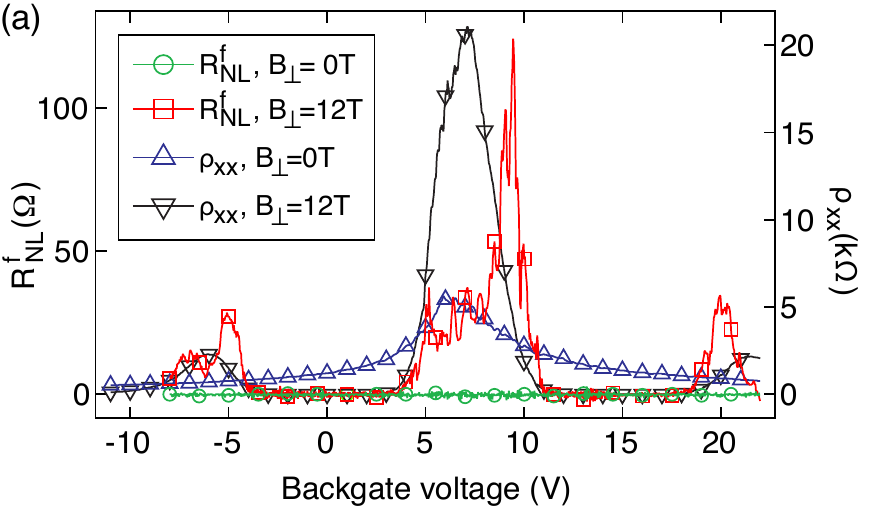}
\includegraphics{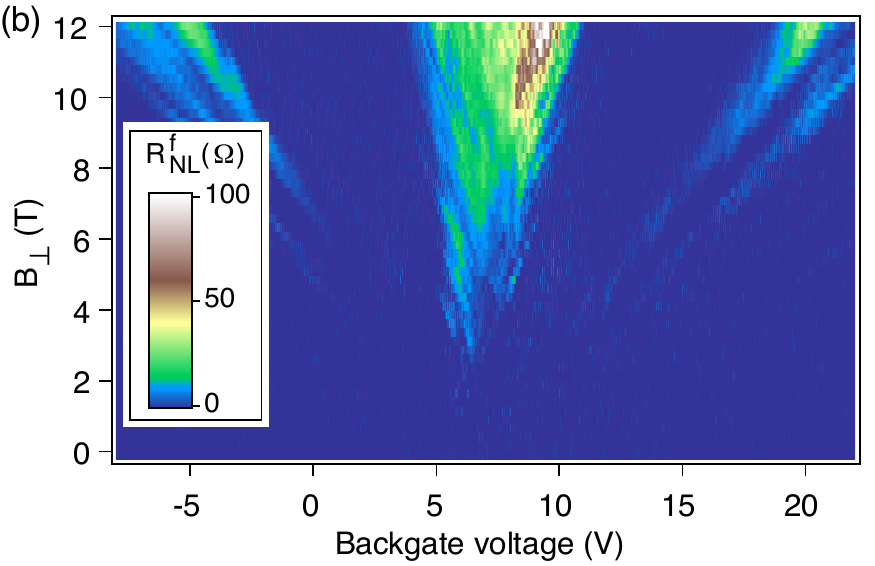}
\caption{\label{fig2} \subref{fig2a}: comparison of  the shapes of $\rho_{xx}$ and $R_{\rm NL}^f$. At zero magnetic field, $R_{\rm NL}^f$ is zero, as expected from classical considerations. At higher magnetic field, there is a strong nonlocal signal, especially near the neutrality point, where $\rho_{xx}$ is maximum as well. The detailed magnetic field dependence of $R_{\rm NL}^f$ \subref{fig2b} shows peaks at $\nu$=$\pm$4 in addition to the dominant central peak. }
\end{figure}

Figure~\ref{fig2a}  presents the basic signature of the effect under investigation in this paper, shown here for sample A.  $R_{\rm NL}^f$ is unmeasurably small at zero magnetic field, as expected due to the vanishing ohmic contribution to nonlocal resistance $\displaystyle\rho_{xx} e^{-\pi a}$ for longitudinal resistivity $\rho_{xx}\lesssim 5k\Omega$ and aspect ratio $a\sim4$ between injector and detector\cite{aba11b}.  When $B_\bot$ is increased to 12T, on the other hand, a large feature in $R_{\rm NL}^f(V_{\rm BG})$  appears at  the charge neutrality point, where filling factor $\nu\equiv nh/eB_\bot=0$, with smaller features at $\nu=\pm4$.  The magnetic field dependence is shown in more detail in Fig.~\ref{fig2b}.  The features at $\nu=\pm4$ correspond to  transitions between quantum Hall plateaus; these are commonly observed in nonlocal measurements \cite{Mce90,aba11b} and attributed to a weak equilibration between bulk and edge state channels \cite{Mce90}.  %As shown in \cite{aba11b}, 
At the neutrality point ($\nu=0$), on the other hand, edge channels are not expected and the signal cannot simply be explained by quantum Hall physics.

\begin{figure}
\subfigure{\label{fig3a}}
\subfigure{\label{fig3b}}
\subfigure{\label{fig3c}}
\includegraphics{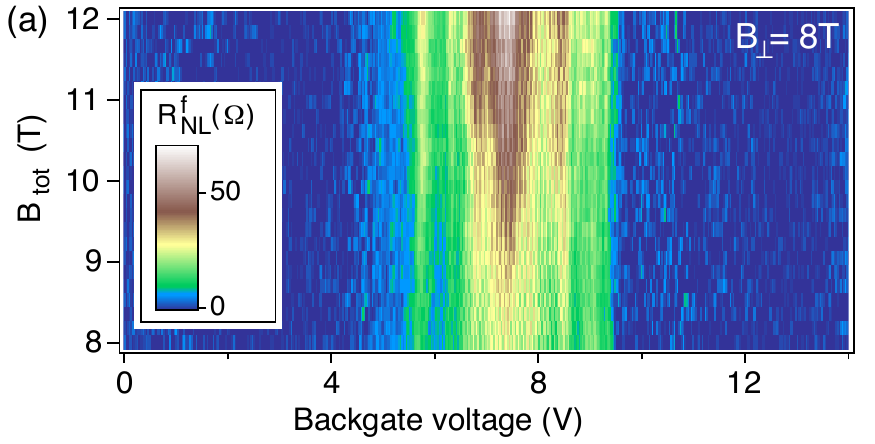}
\includegraphics{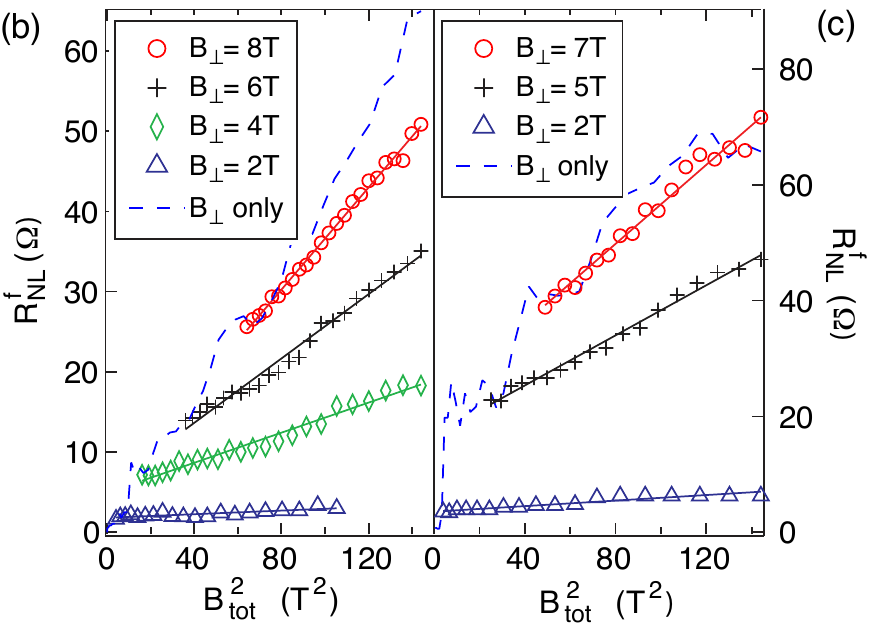}
\caption{\label{fig3} \subref{fig3a}: Dependence of R$_{\rm NL}^f$ on in-plane magnetic field for fixed out-of-plane field, 8 T (sample A).  In-plane magnetic field dependence of R$_{\rm NL}^f$ for samples A \subref{fig3b} and B \subref{fig3c} at 4.2K for several fixed out-of-plane magnetic fields. The lines fits show the quadratic dependence on the total magnetic field: $R_{\rm NL}^f=R_0^f+\beta B_{\rm tot}^2$.  In \subref{fig3b}, $\beta=0.1\pm0.003,0.2\pm0.01,0.32\pm0.01~\Omega/T^2$ for $B_\perp=4T,6T,8T$ respectively; In \subref{fig3c}, $\beta=0.22\pm0.01,0.34\pm0.01~\Omega/T^2$ for $B_\perp=5T,7T$.}
\end{figure}

Spin contributions to the $\nu=0$ feature are quantified by measuring how $R_{\rm NL}^f$ depends on $E_Z$ (that is, on $B_{\rm tot}$) for constant $B_\perp$ (Fig.~3).  As long as $B_\perp$ is held fixed, the measured $\rho_{xx}$  and $\rho_{xy}$ are unaffected by $B_{\rm tot}$ (see supplemental material), whereas $R_{\rm NL}^f$ increases significantly.  This strong dependence on the in-plane component of the magnetic field is a smoking gun of spin-related effects.  As expected for ZSHE, the $\nu=0$ peak in $R_{\rm NL}^f$ increases linearly with $B_{\rm tot}^2$ for many different values of fixed $B_\perp$  (Figs.~\ref{fig3b} and ~\ref{fig3c}).

Looking more closely at the data in Figs.~\ref{fig3b} and \ref{fig3c}, it is clear that the $B_{\rm tot}$ dependence does not extrapolate to zero at $B_{\rm tot}=0$: the data are much better fit by $R_{\rm NL}^f=R_0^f+\beta B_{\rm tot}^2$, compared to the equation $R_{\rm NL}^f=\beta B_{\rm tot}^2$ predicted for ZSHE.   Significant nonlocal signals ($R_0^f$) would be visible even in the absence of Zeeman splitting (though retaining a significant $B_{\perp}$: clearly a physical impossibility).  From this we conclude that a second mechanism contributes to the $\nu=0$ peak in $R_{\rm NL}^f$, which depends on $B_\bot$ but not on $B_{||}$. 

One candidate for nonlocal signals at $n=0$ that depend on $B_\bot$ but not on $B_{||}$ is a valley counterpart to the ZSHE,  predicted in Ref.~\onlinecite{aba11} if the valley degeneracy is lifted.   It was further shown in Ref.~\onlinecite{luk08} that valley splitting is expected from $B_\bot$ (as opposed to $B_{\rm tot}$) since this is an orbital effect. As no other experimental parameter couples to the valley degree of freedom, it is difficult to further test this possibility. On the other hand, one might expect valley currents to be suppressed by the large intervalley scattering rate commonly observed in monolayer graphene.\cite{tikhonenko}

\begin{figure}
\subfigure{\label{fig4a}}
\subfigure{\label{fig4b}}
\subfigure{\label{fig4c}}
\includegraphics{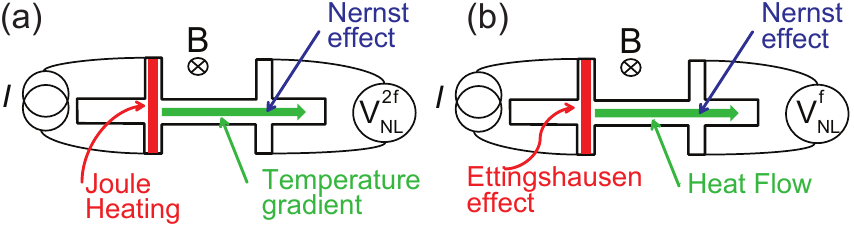}
\includegraphics{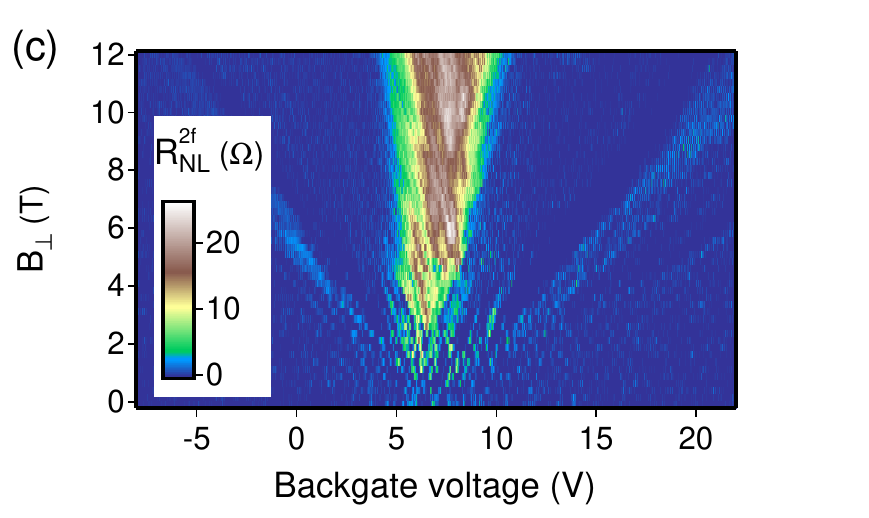}
\caption{\label{fig4} Combinations of thermal effects can give rise to both first harmonic \subref{fig4b} and second harmonic nonlocal signals \subref{fig4a}. \subref{fig4c}: magnetic field dependence of the second harmonic signal (R$_{\rm NL}^{2f}$) measured in sample A at 4.2K.}
\end{figure}

Another candidate for nonlocal voltages at $n=0$ is thermal currents (heat flow) along the graphene. For example, Joule heating at the injector ($\stackrel{.}{Q}_J=I_{B}^2R$) would cause heat to flow into and past the detector region (Fig.~\ref{fig4a}).  The resulting temperature gradient,  $\delta T/\delta x$, along the detector region would give rise to a nonlocal voltage across the detector contacts via the Nernst effect, quantified by the transverse thermopower coefficient $S_{yx}\equiv E_y(\delta T/\delta x)^{-1}\propto V_{\rm NL}(\delta T/\delta x)^{-1}$ (Fig.~\ref{fig4a}).  This temperature gradient is proportional to heating power, quadratic in current, and therefore contributes to the nonlocal voltage only at the second harmonic of the excitation frequency\cite{zue09,che09} (Fig.~\ref{fig4a}).   As a result, Joule heating would not affect the first harmonic data of Figs.~\ref{fig2} and \ref{fig3}.  Fig.~\ref{fig4c} presents the nonlocal second harmonic signal, $R_{\rm NL}^{2f}$, measured at 4.2K with $I_B$=30nA for sample A, for comparison with the first harmonic data from Fig.~\ref{fig2b}.    The complete absence of $B_{||}$ dependence for $R_{\rm NL}^{2f}$ (see supplemental material) is consistent with a thermal origin for this signal.

Another mechanism by which $I_{B}$ can drive heat flow is the Ettingshausen effect (Fig.~\ref{fig4b}), the high magnetic field analog to the Peltier effect just as Nernst is the high-field analog to the Seebeck effect.  Heat flow generated by the Ettingshausen effect, $\stackrel{.}{Q}_E=S_{yx}TI_B$ \cite{del65}  is linear in $I_{B}$, so the resulting $V_{\rm NL}$ would appear at the first harmonic together with  the ZSHE as the offset $R_0^f$ to the linear dependence $\beta B_{\rm tot}^2$ described above.  Is it plausible that the Nernst and Ettingshausen effects together could give rise to an $R_0^f$ as large as that observed in the experiment?  Experimental insight into this question can be gained by using Joule heating---an easily-quantifiable heat source at the injector---to calibrate the Nernst coefficient in our sample.

The following paragraphs offer an order of magnitude calculation to justify the combination of Nernst and Ettingshausen effects as a plausible explanation for the offset $R_0^f\approx5\Omega$ seen in Fig.~\ref{fig2b}.  We start from the approximation that temperature gradient is proportional to heat flow, that is, $\delta T/\delta x=\alpha\stackrel{.}{Q}^{ch}$ where $\stackrel{.}{Q}^{ch}$ is the heat flowing along the channel, past the detector contacts, and $\alpha$ is a proportionality constant that depends on sample geometry and thermal conductivity.   This approximation should be valid as long as $\delta T$ is much less than $T$.  Considering only thermal contributions, the ratio between first (Ettingshausen) and second (Joule) harmonic signals for a given injector current is then $R_{\rm NL}^{2f}$/$R_0^f=\stackrel{.}{Q}_J^{ch}/(S_{yx}TI_B)$, where $\stackrel{.}{Q}_J^{ch}$ is the fraction of Joule heating directed along the channel (heat can also flow away from the detector or to cold injector contacts).

After taking into account the sample geometry and resistivity, we estimate $\stackrel{.}{Q}_J^{ch}\approx3$pW  for sample A near the Dirac point under the conditions B$_{\perp}=$5T, T=4.2K, $I_B$=30nA (see supplementary information for details).  Fig.~\ref{fig4c} gives $R_{\rm NL}^{2f}\approx 20\Omega$ above 5T; if $R_0^f\approx5\Omega$ is to be attributed to Nernst-Ettingshausen we then require $20\Omega/5\Omega=3$pW$/(S_{yx}\cdot4.2$K$\cdot30$nA), that is, $S_{yx}=6\mu$V/K.  This value is remarkably close to estimates of $S_{yx}$ in the literature\cite{zue09, che09}, from which we conclude that the combination of Nernst and Ettingshausen effects can easily explain the offsets observed in Fig.~\ref{fig2}.

%In order to explain the spin-independent part of R$_{\rm NL}^{f}$, the Nernst and Ettingshausen effects should account for the fact that the curves  in Fig.~\ref{fig3} do not extrapolate to zero at zero magnetic field, but rather to a resistance of order 10$\Omega$ when $B_{\perp}\gtrsim 4 T$. In the following paragraphs, we offer an order of magnitude calculation to justify the combination of Nernst and Ettingshausen effects as a plausible explanation.
%, in the following we offer an order estimate the value of the Nernst coefficient required to create such signal. To do so we will determine how this effect should compare to the one we measure at the second harmonic, the origin of which is known.

\begin{figure}
\subfigure{\label{fig5a}}
\subfigure{\label{fig5b}}
\subfigure{\label{fig5c}}
\includegraphics{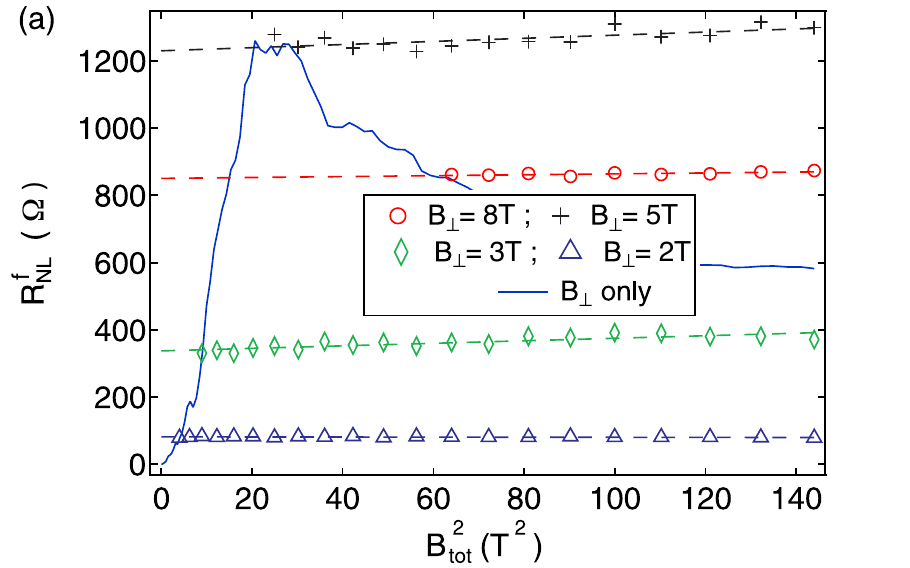}
\includegraphics{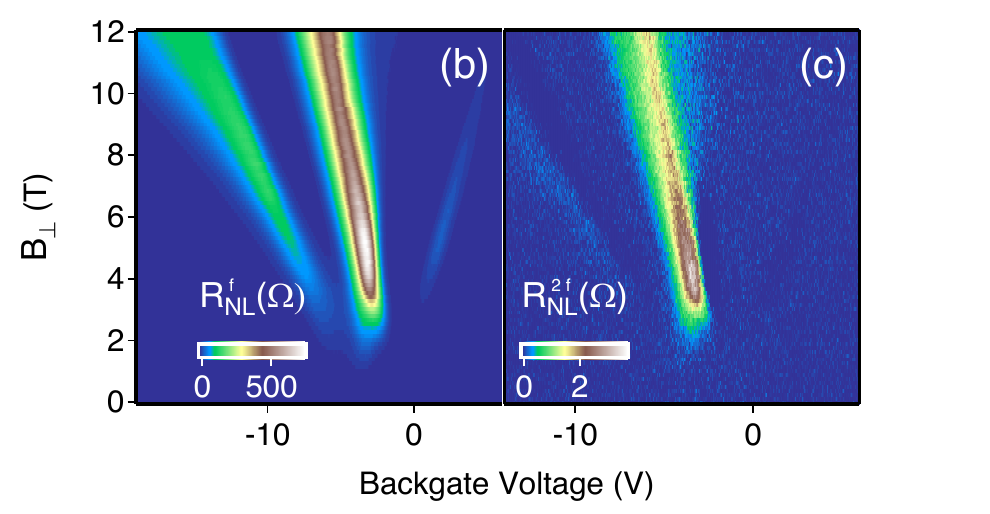}
\caption{\label{fig5} \subref{fig5a}  Nonlocal resistances were much larger in sample C (T=10K), though the in-plane field dependence (markers show data; dashed lines show linear fits) was not stronger than in other samples: $\beta=0.38\pm0.06,0.46\pm0.15,0.14\pm0.05~\Omega/T^2$ for $B_\perp=3T,5T,8T$ respectively.   First harmonic ($R_{\rm NL}^{f}$, \subref{fig5b}) and second harmonic ($R_{\rm NL}^{2f}$, \subref{fig5c}) nonlocal signals (T=77 K) showed similar gate and $B_\bot$ dependence. Further data in supplement.}
\end{figure}

A much stronger thermal signal was observed in sample C (Fig.~\ref{fig5}), where the offset $R_0^{f}$ was two orders of magnitude larger than in samples A and B.  
The extremely large offset created big error bars in the slope $\beta$ of the $B_{||}$ dependence, compared to similar measurements in samples A and B, but the data was consistent with a ZSHE contribution in sample C of similar magnitude as in samples A and B. The comparatively minor effect of $B_{||}$ on the nonlocal signal indicates that the main mechanism behind $R_{\rm NL}^{f}$ in sample C is not related to spin.

On the other hand, the similarity in gate- and field-dependence for the first and second harmonic signal in this sample (Fig.~\ref{fig5b} and \ref{fig5c}) is consistent with a thermal origin. Following the calculation above for sample A, the ratio $R_{\rm NL}^{2f}$/$R_0^{f}$ indicates a much larger coefficient $S_{yx}$ in sample C.  Using values measured at T=10K (data in supplement), we find $R_{\rm NL}^{2f}$/$R_0^{f}$$\approx$0.03 around and B$_{\perp}$=5T for $I_B=100$nA and $\stackrel{.}{Q}_J^{ch}\approx20$pW, giving $S_{yx}\approx$700$\mu$V$\cdot$K$^{-1}$, over two orders of magnitude larger than for samples A and B. (A similar value is obtained from the 77K data in Fig.~\ref{fig5}.)  Although this value of $S_{yx}$ is very large compared with recent reports in graphene, it is on the same order of magnitude as has been reported for graphite \cite{Zhu09}.

%We note that, although contact doping can cause electron-hole asymmetry in general for graphene, we would not expect the contacts to induce the type of magnetic field dependent gate voltage shift of the maximum of the nonlocal signal as seen in Figs.~\ref{fig5b}. This shift could in principle be explained by another (non thermal) source of nonlocality, such as edge transport arising from broken symmetry quantum Hall states. This interpretation would bring other questions, however: why would weak broken symmetry states appear only in sample C, and why would they survive up to relatively high temperatures.%

%To explain the sample to sample variations observed in our measurements, we would need to better understand the mechanisms behind the thermal effects.
This experiment demonstrates that nominally identical samples, with similar electrical and geometric characteristics (mobility, sizes, etc.), display significantly different nonlocal characteristics.  The difference between samples is most notably true with respect to the $B_{||}$-independent component of the signal.  This suggests that thermal transport in graphene may depend very strongly on microscopic sample details that do not influence conventional transport parameters such as mobility.\cite{son12}

%The thermal resistance between the graphene sheet and the SiO$_2$ substrate would strongly modify thermal transport as it would influence the heat going into the channel; this substrate interaction may be very sensitive to  local roughness, which can change from sample to sample, but on the other hand is not a limiting parameter for mobility. The exact nature of the microscopic disorder in the graphene can also play an important role in thermal effects, as predicted by a recent theory . It can change from sample to sample (sample A and B) but also can be different after thermal cycling of a sample (sample A and C).

Finally, we note the non-trivial $B_\bot$ dependence of thermal effects, as seen in Figs.~\ref{fig4c} and \ref{fig5c}.  Unlike the monotonic dependence of ZSHE on B$_{\rm tot}$, the second harmonic signal increased abruptly with $B_\bot$ for low fields, then saturated or even decreased for large B$_\bot$.  Part of the B$_\bot$ dependence may be due to changes in the Nernst coefficient, as reported in Refs.~\onlinecite{zue09,che09}.  It is also likely that the thermal conductivity of the graphene channel depends on $B_\bot$ along with charge conductivity, which could affect the details of the R$_{\rm NL}^{2f}({\rm B}_\bot)$ functional form.  For this reason we compare R$_{\rm NL}^{2f}$ and R$_{\rm NL}^{f}$ only when measured at the same B$_\bot$.

In conclusion, the Zeeman spin Hall effect is an important source of nonlocality in graphene, but in many cases is less strong than nonlocal signals associated with thermal effects.  Spin effects can be clearly distinguished in two out of three samples, while thermomagnetic effects are seen in all, with wide sample-to-sample variations. The study of thermal effects in graphene is a growing field \cite{bet12,bet12b,gra12}; the present study shows that these effects must be taken into account when performing any nonlocal measurement.

\begin{acknowledgments}
 J.R. acknowledges funding from the CIFAR JF Academy and the Max Planck-UBC quantum materials center. M. S acknowledges funding from the Swiss National Science Foundation.
\end{acknowledgments}

%\bibliography{bib_she}

%
% ****** End of file apstemplate.tex ******


\end{document}